\documentclass[3p]{elsarticle}

\usepackage{axodraw}

\usepackage{bm}

\def \in #1 #2 {\int \limits_{#1}^{#2}}

\usepackage{amssymb}
\usepackage{amsmath}
\usepackage{graphicx} 
\usepackage{color}

\usepackage{amsmath}

\usepackage{pstricks}

\def\localinput#1{{
  \renewcommand{\documentclass}[2][dummy]{}
  \renewcommand{\usepackage}[2][dummy]{}
  \renewenvironment{document}{}{}
  \def\jobname{#1}
  \input{#1}
}}

\def\sla#1{\ooalign{\hfil\hspace{-0.1ex}\raise.2ex\hbox{$\not \phantom{#1}$}\hfil\crcr  $#1$}}

\journal{Nuclear Physics B}

\begin{document}



\title{ A Derivation of the Fermi Function in Perturbative Quantum Field Theory }


\author{Akihiro Matsuzaki}
\ead{akihiro@shibaura-it.ac.jp}
\address{Center For Educational Assistance, Shibaura Institute Of Technology,
307 Fukasaku, Minuma-ku, Saitama-shi, Saitama 337-8570 JAPAN 
}

\author{Hidekazu Tanaka}%
\ead{tanakah@rikkyo.ac.jp}
\address{Department of Physics, Rikkyo University,
Nishi-ikebukuro, Toshima-ku Tokyo, Japan, 171
}

\date{\today}

\begin{abstract}

    We postulate that the Fermi function should be derived from the amplitude, not from the solution of the Dirac equation, in the quantum field theory.  
    Then, we obtain the following results. 
    1, We give the amplitude and the width of the neutron beta decay, $n \to p + e^- + \bar \nu_e $ to one loop order. It is carried out by the Feynman parameter integration. 
    2, As the result, we find the terms which can be interpreted as the Fermi function expanded to order $\alpha$.
    3, We also give the same result using complex analysis.
    4, We check that there are no such terms in the similar process, $\bar \nu_e + p \to e^+ + n$.
    5, We perform the Fermi function expanded to order $\alpha^2$ using complex analysis.

\end{abstract}


\maketitle


    The calculation of beta decay rates, for example $n\to p+e^-+\bar \nu_e$, requires the Fermi function. 
    The Fermi function represents the effect that the electron runs through the electromagnetic potential caused by the proton. 
    This function affects the beta spectrum, the decay width, and the lifetime of the parent particle.

    The Fermi function has been derived so far as the solution of the Dirac equation in electromagnetic potential \cite{ZPhys88-161} \cite{AmJP36-1150} \cite{ProcRoySoc133-381} \cite{NPA273-301}. 
    In beta decay, the decay itself is caused by the weak interaction and treated as the intermediate state, which is represented as the amplitude.
    The final state particles are in electromagnetic potential.
    This effect is factorized as the Fermi function \cite{NPA337-474}.
    For the non-relativistic limit in neutron beta decay, it takes the form  
\begin{align} \begin{split}\label{Eq-Fermi-Function}
F_{NR}
=\ \frac{2\pi\alpha/v}{1-e^{-2\pi\alpha/v}},
\end{split} \end{align}
    where $\alpha$ is the fine structure constant and $v$ is the electron velocity in the neutron rest frame. 
    Calculating the decay width, the Fermi function is multiplied by the absolute square of the amplitude, and integrated over the phase space.
    This is the same for the loop amplitude.
    For the sake of simplicity, in this paper, we set the parent and daughter nucleons as neutron and proton, respectively.

    From a quantum field theoretical point of view, this potential effect is also the interaction of intermediate state.
    Generally, the created particles should be considered as the asymptotic fields in the far future. 
    The asymptotic fields are free from the interactions between the created particles.
    The electromagnetic interaction should be derived from the loop diagrams.
    Furthermore, this effect should appear in perturbation theory as
\begin{align} \begin{split}\label{Eq-Expanded-Fermi-Function}
F_{NR}
=1+\frac{\pi\alpha}{v}+\frac{\pi^2\alpha^2}{3v^2}-\frac{\pi^4\alpha^4}{45v^4}+\cdots
\end{split} \end{align}
    order by order with respect to $\alpha$.
    We note that $F_{NR}$ is actually expanded with respect to $\alpha/v$ rather than $\alpha$.



    The tree level beta decay diagram is
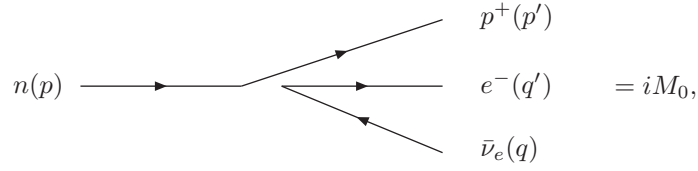
\begin{center}
\fcolorbox{white}{white}{
  \begin{picture}(180,61) (60,-44)
    \SetWidth{0.5}
    \SetColor{Black}
    \SetWidth{0.5}
    \ArrowLine(60,-14)(120,-14)
    \ArrowLine(135,-14)(195,-14)
    \ArrowLine(120,-14)(195,11)
    \ArrowLine(195,-39)(135,-14)
    \Text(260,-19)[lb]{\Black{$=iM_0$,}}
    \Text(35,-19)[lb]{\Black{$n(p)$}}
    \Text(210,-19)[lb]{\Black{$e^-(q')$}}
    \Text(210,-43)[lb]{\Black{$\bar{\nu}_e(q)$}}
    \Text(210,7)[lb]{\Black{$p^+(p')$}}
  \end{picture}
}
\end{center}
    where the parameters in each parenthesis represent their momenta. 
    Also, the one loop diagrams are
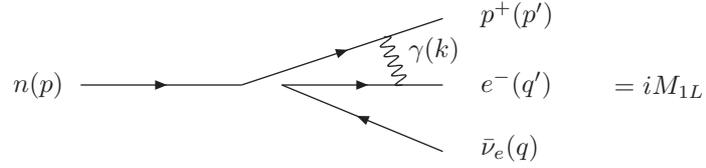
\begin{center}
\fcolorbox{white}{white}{
  \begin{picture}(210,61) (30,-44) 
    \SetWidth{0.5}
    \SetColor{Black}
    \ArrowLine(60,-14)(120,-14)
    \ArrowLine(135,-14)(195,-14)
    \ArrowLine(120,-14)(195,11)
    \ArrowLine(195,-39)(135,-14)
    \Text(260,-19)[lb]{\Black{$=iM_{1L}$}}
    \Text(35,-19)[lb]{\Black{$n(p)$}}
    \Text(210,-19)[lb]{\Black{$e^-(q')$}}
    \Text(210,-43)[lb]{\Black{$\bar{\nu}_e(q)$}}
    \Text(210,7)[lb]{\Black{$p^+(p')$}}
    \Text(183,-6)[lb]{\Black{$\gamma(k)$}}
    \Photon(180,-14)(173,4){2.5}{5}
  \end{picture}
}
\begin{align}  \label{One-Loop-Diagram}
\end{align}
\end{center}
    and the field strength renormalization.
    According to Ref. \cite{PLB598-67}, the one loop amplitude is divided into three parts as $iM_{1L}=iM_{1}+iM_{2}+iM_{3}$.
    We calculate only $iM_1+iM_2$, which do not depends on $\sigma^{\mu\nu}k_\nu$ in the numerator of proton propagator, where $\sigma^{\mu\nu}=i(\gamma^\mu\gamma^\nu-\gamma^\nu\gamma^\mu)/2$.
    To cancel the infrared divergence, we consider the sum of two bremsstrahlung diagrams, $iM_b$.
    
    The detailed calculations are given in Appendix \ref{App1}, and the result is 
\begin{align} \begin{split}\label{Eq-width1}
d\Gamma
=&d\Gamma_3+
\frac{1}{\pi}G_F^2\frac{d^3q'}{(2\pi)^3}(1+3C^2)k_M^2\biggl[1+\frac{\alpha}{2\pi}\biggl\{\frac{2\pi^2}{v}+3\log\frac{m_p}{m_e}-\frac{1}{2}
-\frac{4}{v}Li\left(\frac{2v}{1+v}\right)\\
&+4\left(\frac{1}{v}\mathrm{Tanh}^{-1}v-1\right)\left(\frac{k_M}{3E_e}-\frac{3}{2}+\log\frac{2k_M}{m_e}\right)
+\frac{1}{v}\mathrm{Tanh}^{-1}v\left(2(1+v^2)+\frac{k_M^2}{6E_e^2}-4\mathrm{Tanh}^{-1}v\right)\biggr\}\biggr],
\end{split} \end{align}
    where $G_F$, $m_n$, $m_p$, $m_e$, and $E_e$ are the Fermi constant, neutron mass, proton mass, electron mass, and electron energy, respectively; 
    $k_M=m_n-m_p-E_e$ represents the maximum photon energy for given $E_e$;
    $C$ represents the Gamow-Teller coupling constant relative to the Fermi transition;
    $Li(x)$ is the Spence function defined as
\begin{align} \begin{split}
Li(x)=-\int^{x}_{0} dz \frac{\log(1-z)}{z}.
\end{split} \end{align}    
    We set $m_n\simeq m_p \gg E_e,m_e$ and the neutrino is massless.
    $d\Gamma_3$ is a part depending on $iM_3$, which does not depend on $v$ \cite{PLB598-67}.
    We do not calculate $d\Gamma_3$ in this paper.
    The electron velocity is represented as $v=|\bm{q}'|/q_0'$ in the neutron rest frame.
    The first term in the curly brackets of Eq. (\ref{Eq-width1}) is inversely proportional to $v$. 
    This term is interpreted as the one corresponds to order $\alpha$ Fermi function expressed in Eq. (\ref{Eq-Expanded-Fermi-Function}).  
    We note here, the Eq. (\ref{Eq-width1}) differs $1/v$ term and the constant from Ref. \cite{PLB598-67}.

    According to Appendix \ref{App1}, the amplitude is approximately written as
\begin{align} \begin{split}\label{Eq-1loop-amp}
iM_0+iM_{1}
\ni
iM_0(1+\frac{\alpha}{4\pi}I_{5a})
\ni
&iM_0\left\{1+\frac{\pi\alpha}{2v}+i\frac{\alpha}{2v}\log\left(\frac{4m_e^2}{\mu^2}\frac{v^2}{1-v^2}\right)\right\}
\end{split} \end{align}
    for $v\ll1$, where $\mu$ is the photon mass introduced to regulate the infrared divergence.
    The last term does not affect the one loop width, but affects the two loop one as explained later. 

    Here, we extract Eq. (\ref{Eq-1loop-amp}) using complex analysis.
    According to Appendix \ref{App1}, $I_{5a}$ contains this term.
    $I_{5a}$ originates from one loop diagram in Eq. (\ref{One-Loop-Diagram}), not from the bremsstrahlung or field strength renormalization terms.
    Furthermore, $I_{5a}$ does not contain $k$ in its numerator.
    We explain this reason later. 
    Therefore, we start from 
\begin{align} \begin{split}\label{Eq-1l-amp-2}
iM_1\ni& \int \frac{d^4 k}{(2\pi)^4}\ \frac{-4e^2 M_0 p'\cdot q'}{(p'-k)^2-m_p^2+i\epsilon}\ \frac{1}{(q'+k)^2-m_e^2+i\epsilon}\ \frac{1}{k^2-\mu^2+i\epsilon},
\end{split} \end{align}
    where $i\epsilon$ represents the Feynman prescription; $e$ is the electromagnetic coupling constant.
     
    Next, we integrate Eq. (\ref{Eq-1l-amp-2}) with respect to $k_0$, where the contour integral is taken clockwise.
    The locations of poles are  
$k_0=p_0'+\sqrt{{p_0'}^2+\bm{k}^2-i\epsilon}$, $k_0=-q_0' + \sqrt{{q_0'}^2+\bm{k}^2-2\bm{q}'\cdot\bm{k}-i\epsilon}$, and $k_0=\sqrt{\bm{k}^2+\mu^2-i\epsilon}$.
    Since we focus on the terms proportional to $\alpha/v$, which dominate for $v=|\bm{q}'|/q'_0\ll 1$, we set $p_0'\gg q_0'\gg|\bm{q}'|$.
    The contributions which converge for $v\to0$ can be ignored.
    For $|\bm{k}| \hspace{0.3em}\raisebox{0.4ex}{$>$}\hspace{-0.75em}\raisebox{-.7ex}{$\sim$}\hspace{0.3em}  q_0'$, the integral over $|\bm{k}|$ converges even if we set $v\to0$.
    Hence, we set $q_0'\gg|\bm{k}|$, and the locations of poles are $k_0\simeq2p_0'$, $k_0\simeq \frac{1}{2q_0'}(\bm{k}^2-2\bm{q}'\cdot\bm{k}-i\epsilon)$, and $k_0\simeq \sqrt{\bm{k}^2+\mu^2-i\epsilon}$.
    Writing only the contribution from the second pole explicitly, 
\begin{align} \begin{split}
iM_1\ni& 2e^2 q_0' iM_0\int \frac{d^3 k}{(2\pi)^3}\frac{1}{\bm{k}^2+2\bm{q}'\cdot\bm{k}-i\epsilon}\frac{1}{\bm{k}^2+\mu^2-i\epsilon}+\cdots,
\end{split} \end{align}
    where the part, $"+\cdots"$ represents the contribution from other poles.
    
    Defining $\bm{x}=\bm{k}/|\bm{q}'|$, $|\bm{x}|=x$, $\cos\theta=\bm{q}'\cdot\bm{k}/(|\bm{q}'||\bm{k}|)$, and $\bar{\mu}=\mu/|\bm{q}'|$, the integrand becomes dimensionless as
\begin{align} \begin{split}\label{Eq-iM1-x-1}
iM_1\ni& \frac{e^2 iM_0}{2\pi^2}\frac{q_0'}{|\bm{q}'|} \int \frac{d\cos\theta d x}{x^2+2x\cos\theta-i\epsilon}\frac{x^2}{x^2+\bar{\mu}^2-i\epsilon}+\cdots
\ni\frac{\alpha iM_0}{\pi v} \int_0^\infty x dx\frac{\log(x^2+2x-i\epsilon)-\log(x^2-2x-i\epsilon)}{x^2+\bar{\mu}^2-i\epsilon}.\\
\end{split} \end{align}
    In the right hand side, the part $"+\cdots"$ is ignored since these terms are not proportional to the factor $1/v$. 
    Generally, the dimensionless integrand does not contain the factor $v$.
    Then, it can be ignored that the terms which do not have the factor $1/v$ as a coefficient of the integral.
    Here, if the numerator of the integrand have the term which contains $k$, it gives additional factor $|\bm{q}'|$ when we take the integrand dimensionless.
    Then, such a term cannot be the candidate of the Fermi function.    

    We use complex analysis to the $x$ integral. 
    Since the integrand in Eq. (\ref{Eq-iM1-x-1}) is the even function, the amplitude can be written as 
\begin{align} \begin{split}\label{Eq-1Loop-Amp-a}
iM_1\ni&\frac{\alpha}{2\pi v} iM_0\int_{-\infty}^\infty \frac{x dx}{x^2+\bar{\mu}^2-i\epsilon}\bigl[\log(x^2+2x-i\epsilon)-\log(x^2-2x-i\epsilon)\bigr].
\end{split} \end{align}
    To apply the residue theorem, we integrate Eq. (\ref{Eq-1Loop-Amp-a}) by parts to form
\begin{align} \begin{split}
iM_1\ni&\frac{\alpha}{\pi v} iM_0\int_{-\infty}^\infty   dx 
\frac{x^2}{(x^2+2x-i\epsilon)(x^2-2x-i\epsilon)}\log(x^2+\bar{\mu}^2-i\epsilon).
\end{split} \end{align}
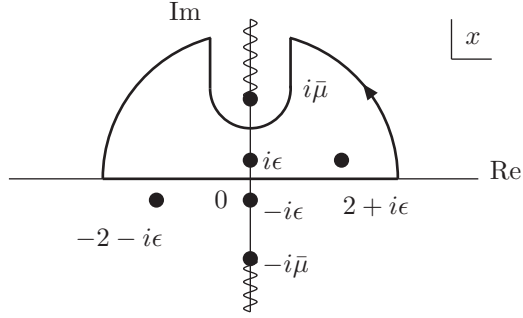
\begin{figure}[htbp]
\begin{center}
\fcolorbox{white}{white}{
  \begin{picture}(210,126) (15,11)
    \SetWidth{0.5}
    \SetColor{Black}
    \Line(105,121)(105,11)
    \Text(195,61)[lb]{\Black{$\mathrm{Re}$}}
    \Text(75,121)[lb]{\Black{$\mathrm{Im}$}}
    \Line(180,121)(180,106)
    \Line(180,106)(195,106)
    \Photon(105,91)(105,121){2.5}{5}
    \Vertex(105,31){2.83}
    \Vertex(105,91){2.83}
    \Photon(105,31)(105,11){2.5}{4}
    \Text(92,50)[lb]{\Black{$0$}}
    \Line(15,61)(190,61)
    \SetWidth{1.0}
    \CArc(105,61)(55,105.5,180)
    \Line(50,61)(160,61)
    \ArrowArc(105,61)(55,0,74.5)
    \Line(120,114)(120,95)
    \Line(90,114)(90,95)
    \CArc(105.02,95.02)(15.02,-179.94,3.76)
    \Text(125,90)[lb]{\Black{$i\bar{\mu}$}}
    \Text(110,24)[lb]{\Black{$-i\bar{\mu}$}}
    \Text(186,112)[lb]{\Black{$x$}}
    \SetWidth{0.5}
    \Vertex(139,68){2.83}
    \Vertex(70,53){2.83}
    \Vertex(105,68){2.83}
    \Vertex(105,53){2.83}
    \Text(110,65)[lb]{\Black{$i\epsilon$}}
    \Text(110,45)[lb]{\Black{$-i\epsilon$}}
    \Text(140,47)[lb]{\Black{$2+i\epsilon$}}
    \Text(40,35)[lb]{\Black{$-2-i\epsilon$}}
  \end{picture}
}
  \caption{$x$ contour and the location of the poles}
\label{1loop-x-contour}
\end{center}
\end{figure}
    Applying the residue theorem, the amplitude becomes 
\begin{align} \begin{split}\label{Eq-Amp-1L-b}
iM_1\ni&iM_0\left(\frac{\pi\alpha}{2v}-i\frac{\alpha}{v}\log\frac{\bar{\mu}}{2}\right)
=iM_0\left\{\frac{\pi\alpha}{2v}+i\frac{\alpha}{2v}\log\left(\frac{4m_e^2}{\mu^2}\frac{v^2}{1-v^2}\right)\right\},
\end{split} \end{align}
    where the contour is depicted in Fig. \ref{1loop-x-contour}.

    This result is consistent with Eq. (\ref{Eq-1loop-amp}) not only the real part but also the imaginary part in the curly brackets.


    More than one charged particles does not exist at the same time in the scattering process $\bar \nu_e + p \to e^+ + n$.
    Therefore, the electron and proton are not affected by the electromagnetic potential and the amplitude should not contain the $\alpha/v$ terms which can be interpreted as a part of the Fermi function.
    Here, we verify it. 

    According to Ref. \cite{APPB35-1687}, the diagram is    
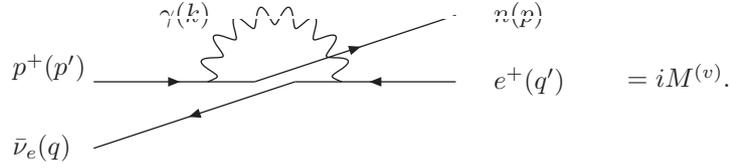
\begin{center}
\fcolorbox{white}{white}{
  \begin{picture}(210,51) (30,-29)
    \SetWidth{0.5}
    \SetColor{Black}
    \ArrowLine(60,1)(120,1)%
    \ArrowLine(195,1)(135,1)%
    \ArrowLine(120,1)(195,26)
    \ArrowLine(135,1)(60,-24)
    \Text(260,-1)[lb]{\normalsize{\Black{$=iM^{(v)}.$}}}
    \Text(30,-29)[lb]{\normalsize{\Black{$\bar{\nu}_e(q)$}}}
    \Text(210,-4)[lb]{\normalsize{\Black{$e^+(q')$}}}
    \Text(30,1)[lb]{\normalsize{\Black{$p^+(p')$}}}
    \Text(210,21)[lb]{\normalsize{\Black{$n(p)$}}}
    \PhotonArc(127.5,1)(25,0,180){-3.5}{8.5}
    \Text(85,21)[lb]{\normalsize{\Black{$\gamma(k)$}}}
  \end{picture}
}
\end{center}
    Extracting the related terms by the similar manner as the beta decay, 
\begin{align} \begin{split}
iM^{(v)}
\ni& \int \frac{d^4 k}{(2\pi)^4}\frac{4e^2 p'\cdot q' M_0' }{(p'-k)^2-m_p^2+i\epsilon}\frac{1}{(q'-k)^2-m_e^2+i\epsilon}\frac{1}{k^2-\mu^2+i\epsilon}.
\end{split} \end{align}
    We integrate with respect to $k_0$ along the counter-clockwise.
    After the same approximation with the beta decay, the poles we focus on are 
$k_0\simeq-\frac{1}{2p_0'}(\bm{k}^2-i\epsilon)$ and $k_0=-\frac{1}{2q_0'}(\bm{k}^2-2\bm{q}'\cdot\bm{k}-i\epsilon)$.
    Each residue is the same value with the opposite sign.
    Then, the amplitude becomes
\begin{align} \begin{split}
iM^{(v)}
\ni& 2e^2 q_0' iM_0'\int \frac{d^3 k}{(2\pi)^3}\frac{1}{\bm{k}^2+2\bm{q}'\cdot\bm{k}-i\epsilon}\frac{1}{\bm{k}^2+\mu^2-i\epsilon} 
-  2e^2 q_0' iM_0'\int \frac{d^3 k}{(2\pi)^3}\frac{1}{\bm{k}^2+2\bm{q}'\cdot\bm{k}-i\epsilon}\frac{1}{\bm{k}^2+\mu^2-i\epsilon}
=0,\nonumber
  \end{split} \end{align}
    where $iM_0'$ is the tree level amplitude.

    As a result, this scattering process does not have $\alpha/v$ term, which can be interpreted as the part of the Fermi function.
    This is because the signs on $k$ in the electron and proton propagators are the same. 
    It is just equivalent to the two charged particles do not exist at the same time.
    This supports that the $\alpha/v$ terms in beta decay correspond to the potential effect.


    Two loop ladder diagram is
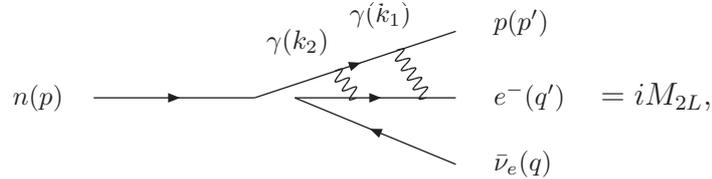
\begin{center}
\fcolorbox{white}{white}{
  \begin{picture}(210,61) (30,-44)
    \SetWidth{0.5}
    \SetColor{Black}
    \ArrowLine(60,-14)(120,-14)
    \ArrowLine(120,-14)(195,11)
    \Text(210,9)[lb]{\normalsize{\Black{$p(p')$}}}
    \Text(30,-19)[lb]{\normalsize{\Black{$n(p)$}}}
    \ArrowLine(135,-14)(195,-14)
    \ArrowLine(195,-39)(135,-14)
    \Text(250,-19)[lb]{\normalsize{\large{\Black{$=iM_{2L}$,}}}}
    \Text(210,-19)[lb]{\normalsize{\Black{$e^-(q')$}}}
    \Text(210,-44)[lb]{\normalsize{\Black{$\bar{\nu}_e(q)$}}}
    \Photon(158,-14)(150,-3.9){2.5}{3}
    \Photon(184,-14)(173,3.9){2.5}{5}
    \Text(125,2)[lb]{\Black{$\gamma(k_2)$}}
    \Text(156,12)[lb]{\Black{$\gamma(k_1)$}}
  \end{picture}
}
\end{center}
    where $k_1$ and $k_2$ are the photon momenta, respectively. 
    There are some other two loop diagrams.
    However, we are only interested in the term proportional to $(\alpha/v)^{2}$.
    This term originates only from $iM_{2L}$. 
    After the similar calculation to the one loop amplitude, the amplitude contains 
\begin{align} \begin{split}
iM_{2L}
\ni
&-iM_0\frac{\alpha^2  }{v^2}
\left(-\frac{1}{24}\pi^2+\frac{1}{2}\log^2\frac{\bar{\mu}}{2}\right).
\end{split} \end{align}

    Summing this equation, Eq. (\ref{Eq-Amp-1L-b}), and the tree level amplitude, we give
\begin{align} \begin{split}\label{Eq-Amp-2L-1}
iM_0+iM_{1L}+iM_{2L}
\ni
&iM_0\left\{1+\frac{\pi\alpha}{2v}-\frac{i\alpha}{v}\log\frac{\bar{\mu}}{2}
-\frac{\alpha^2  }{v^2}
\left(-\frac{1}{24}\pi^2+\frac{1}{2}\log^2\frac{\bar{\mu}}{2}\right)\right\}.
\end{split} \end{align}
    The absolute square of them is 
\begin{align} \begin{split}
\left|iM_0+iM_{1L}+iM_{2L}\right|^2
\ni
&\left|iM_0\right|^2
\left(1+\frac{\pi\alpha}{v}+\frac{\pi^2\alpha^2}{3v^2}\right)
+\mathcal{O}(\alpha^3).
\end{split} \end{align}
    The logarithmic terms in Eq. (\ref{Eq-Amp-2L-1}) are canceled. 
    For $v\ll1$, the decay width has the form 
\begin{align} \begin{split}
d\Gamma-d\Gamma_3\propto
1+\frac{\pi\alpha}{v}+\frac{\pi^2\alpha^2}{3v^2}.
\end{split} \end{align}
    This is consistent with the Fermi function up to order $\alpha^2$


    We conclude the main results as follows. 
\begin{enumerate}
  \item We reviewed the one loop beta decay amplitude to find the terms proportional to $\alpha/v$. It can be interpreted as the part of the Fermi function.
  \item The scattering process $\bar \nu_e + p \to e^+ + n$ does not have such terms.
  \item We give the result that the two loop beta decay amplitude has the terms proportional to $(\alpha/v)^2$. These are consistent with the expanded Fermi function up to order $\alpha^2$.
\end{enumerate}

    These results suggest that the potential effect named the Fermi function should be considered as a part of the amplitude.

    To confirm our conclusion, it is necessary to carry out the Fermi function to higher order. 
    If the systematic calculation will be carried out, we can sum up all the order of contributions.

    For $\alpha/v \hspace{0.3em}\raisebox{0.4ex}{$>$}\hspace{-0.75em}\raisebox{-.7ex}{$\sim$}\hspace{0.3em} 1$, the perturbation up to the finite order does not work.
    We must sum up all order of $\alpha/v$.
    The result should become the full Fermi function written in Eq. (\ref{Eq-Fermi-Function}).
    Then, we propose the decay width to form
\begin{align} \begin{split}
d\Gamma-d\Gamma_3
=&
\frac{1}{\pi}G_F^2\frac{d^3q'}{(2\pi)^3}(1+3C^2)k_M^2\biggl[ \frac{2\pi\alpha/v}{1-e^{-2\pi\alpha/v}}+\frac{\alpha}{2\pi}\biggl\{3\log\frac{m_p}{m_e}-\frac{1}{2}
-\frac{4}{v}Li\left(\frac{2v}{1+v}\right)\\
&+4\left(\frac{1}{v}\mathrm{Tanh}^{-1}v-1\right)\left(\frac{k_M}{3E_e}-\frac{3}{2}+\log\frac{2k_M}{m_e}\right)
+\frac{1}{v}\mathrm{Tanh}^{-1}v\left(2(1+v^2)+\frac{k_M^2}{6E_e^2}-4\mathrm{Tanh}^{-1}v\right)\biggr\}\biggr]. \nonumber
\end{split} \end{align}

    In two loop calculation, we were only interested in the terms proportional to $(\alpha/v)^2$.
    However, the terms proportional to $\alpha^2/v$ may exist.
    These terms also affect the decay width.
    We should consider them for the higher order calculation.      

    Our result does not affect the practical use except for $v \hspace{0.3em}\raisebox{0.4ex}{$<$}\hspace{-0.75em}\raisebox{-.7ex}{$\sim$}\hspace{0.3em} \alpha$.
    For instance, this result only slightly affects the Kurie plot \cite{PLB598-188} and Ref. \cite{PTP123-1003}.
    However, the theoretical calculation of the nuclear lifetime is changed.

    This study can be applied to beta decay of other nuclear species by exchanging $\alpha\to Z \alpha$, where $Z$ is the atomic number of the parent particle.
    The loop diagrams which contain the photon propagator between parent nuclear and the daughter particles do not give the $\alpha/v$ term as explained in the calculation of the scattering process. 
    Also, we confirmed that the $(\alpha/v)^2$ term does not appear in the corresponding two loop diagrams.
    Our study is more important for the beta decay of nuclear which have larger $Z$.

\appendix

\section{the Detail of One Loop Calculation}\label{App1}

    The tree level amplitude is
\begin{align} \begin{split}
iM_0=-\frac{iG_F}{\sqrt{2}}\bar{u}(p')(1-C\gamma^5)u(p)\bar{u}(q')(1-\gamma^5)v(q).
\end{split} \end{align}
    According to Ref. \cite{PLB598-67}, the one loop amplitude can be separate in three parts as 
\begin{align} \begin{split}
iM_{\mathrm{1L}}=iM_1+iM_2+iM_3.
\end{split} \end{align}
    $iM_1$ picks up the factors $(2q'+k)^\mu$ from the electron propagator and $(2p'-k)_\mu$ from the proton propagator.
    $iM_2$ picks up the factors $\sigma^{\mu \nu}k_\nu$ from the electron propagator and $(2p'-k)_\mu$ from the proton propagator.
    $iM_3$ picks up the remaining factors.


    The tree level width is 
\begin{align} \begin{split}
d\Gamma_0
=&\frac{1}{8\pi m_n m_p}\frac{d^3q'}{(2\pi)^3}\frac{1}{2E_e}\frac{1}{2}\sum_\varepsilon |iM_0|^2\frac{k_M^2}{E_\nu},\ \ \ \ \ \ 
\sum_\varepsilon\left|iM_0\right|^2
=32G_F^2m_n m_p E_e E_\nu(1+3C^2),
\end{split} \end{align}
    where $E_\nu$ and $m_n$ are the neutrino energy and the neutron mass, respectively;
    $\sum_\varepsilon$ represents the spin sums.

    The one loop width is
\begin{align} \begin{split}
d\Gamma
=&d\Gamma_b+\frac{1}{2m_n} \left(\frac{d^3p'}{(2\pi)^3}\frac{1}{2E_{p'}}\right)\left( \frac{d^3q'}{(2\pi)^3}\frac{1}{2E_{q'}}\right)\left( \frac{d^3q}{(2\pi)^3}\frac{1}{2E_{q}}\right)\\
&\times
\frac{1}{2}\sum_\varepsilon \left|iM_0+\frac{1}{2}(\delta Z_e+\delta Z_p)iM_0+iM_{1L}\right|^2\delta^{(4)}(p-p'-q-q'),\\
d\Gamma_b=&\frac{1}{2m_n}\left( \frac{d^3p'}{(2\pi)^3}\frac{1}{2E_{p'}}\right)\left( \frac{d^3q'}{(2\pi)^3}\frac{1}{2E_{q'}}\right)\left( \frac{d^3q}{(2\pi)^3}\frac{1}{2E_{q}}\right)\left( \frac{d^3k}{(2\pi)^3}\frac{1}{2E_{\gamma}}\right)
 \frac{1}{2}\sum_\varepsilon |iM_b|^2\delta^{(4)}(p-p'-q-q'-k),  \nonumber
\end{split} \end{align}
    where $\delta Z_e$ and $\delta Z_p$ are the one loop electron and proton field strength renormalization, respectively;
    $d\Gamma_b$ is the term originated from the bremsstrahlung; $d\Gamma_3$ is the term originated from $iM_3$. 

%
%
%
    Here, $iM_2$ is written as 
\begin{align} \begin{split}
iM_2=&
\frac{2iJ}{(4\pi)^2}\left[\frac{1}{2c q'^2}\biggl\{\frac{b-c}{1+b-c}\log(b-c)-\frac{b+c}{1+b+c}\log(b+c)\biggr\}-\frac{i\pi}{(p'+q')^2}\right],
\\J=&
-\frac{e^2G}{\sqrt{2}}
\bar u(p')\gamma^\nu(1-C_A \gamma_5)u(p)
\bar u(q')(p'\cdot q'-m_e \sla p')\gamma_\nu(1-\gamma_5)v(q),
\end{split} \end{align}
    Then, the cross term between $iM_0$ and $iM_2$ is 
\begin{align} \begin{split}
\sum_\varepsilon iM_0 \left(iM_2\right)^*
&\simeq
\frac{\alpha }{4\pi} \sum_\varepsilon |iM_0|^2
v\left(\log\frac{1+v}{1-v}-\frac{i\pi}{m_p^2}\right).
\end{split} \end{align}
    Then, the decay width takes the form
\begin{align} \begin{split}\label{25}
d\Gamma
=&d\Gamma_0\left\{1+\frac{\alpha}{2\pi}\mathrm{Re}\left(\sum_{i=1}^{6}I_i+v\log\frac{1+v}{1-v}\right)\right\}+d\Gamma_3+d\Gamma_b,
\end{split} \end{align}
    where $I_i$ are defined in Appendix \ref{App-Ii}.

\subsection{$d\Gamma_b$}\label{App-dGamma_b}

    The bremsstrahlung amplitude is approximately written as \cite{PR113-1652}
\begin{align} \begin{split}
iM_b\simeq eiM_0\biggl(\frac{q'\cdot \epsilon(k)}{q'\cdot k+i\epsilon}-\frac{p'\cdot \epsilon(k)}{p'\cdot k+i\epsilon}\biggr)
\end{split} \end{align}
    for small $k$, where $\epsilon_\mu(k)$ in the numerator represents the polarization vector of the external photon.

    The absolute square of this amplitude is 
\begin{align} \begin{split}
\sum_\varepsilon |iM_b|^2\simeq \sum_\varepsilon |iM_0|^2 \frac{e^2}{E_e}\biggl[\frac{1}{E_e(1-v\beta w)}+\frac{E_e+k_0}{k_0^2}\frac{v^2(1-\beta^2 w^2)}{(1-v\beta w)^2}\biggr],
\end{split} \end{align}
    where $\beta=|\bm{k}|/k_0$ and $w=\bm{k}\cdot\bm{q}'/(|\bm{k}||\bm{q}'|)$.
    Here, we define $k=|\bm{k}|$, and 
\begin{align} \begin{split}
I_b \equiv& \int^{1}_{-1} dw \int^{k_M}_{0} dk \frac{k^2}{k_0}\biggl[\frac{1}{E_e^2(1-v\beta w)}+\frac{E_e+k_0}{E_e}\frac{v^2(1-\beta^2 w^2)}{k_0^2(1-v\beta w)^2}\biggr]\left(1-\frac{k_0}{k_M}\right)^2\\
=&2\biggl[2+\frac{k_M^2}{12E_e^2}-\frac{1}{v}Li(\frac{2v}{1+v})-\frac{1}{v}(\mathrm{Tanh}^{-1}v)^2
+(2-\frac{2k_{M}}{3E_e}-\frac{k_M^2}{12E_e^2}-2\log\frac{2k_{M}}{\mu})(1-\frac{1}{v}\mathrm{Tanh}^{-1}v)\biggr],
\end{split} \end{align}
    where $\mu=\sqrt{k_0^2-|\bm{k}|^2}$ is the photon mass. 
    Then, the bremsstrahlung part is
\begin{align} \begin{split}
d\Gamma_b
=&\frac{1}{8\pi m_n m_p}\frac{d^3q'}{(2\pi)^3}\frac{1}{2E_e}\frac{1}{2}\sum_\varepsilon |iM_0|^2\frac{k_M^2}{E_\nu} \times \frac{\alpha}{2\pi} I_b.
\end{split} \end{align}
    Therefore, the Eq. (\ref{25}) becomes 
\begin{align} \begin{split}
d\Gamma-d\Gamma_3
=&d\Gamma_0 \left[1+\frac{\alpha}{2\pi}\left\{\mathrm{Re}\left(\sum_{i=1}^{6}I_i+v\log\frac{1+v}{1-v}\right)+ I_b \right\}\right].
\end{split} \end{align}
    According to Appendix \ref{App-Ii}, the one loop width finally takes the form
\begin{align} \begin{split}
d\Gamma-d\Gamma_3
=&
\frac{1}{\pi}G_F^2\frac{d^3q'}{(2\pi)^3}(1+3C^2)k_M^2\biggl[1+\frac{\alpha}{2\pi}\biggl\{\frac{2\pi^2}{v}+3\log\frac{m_p}{m_e}-\frac{1}{2}-\frac{4}{v}Li\left(\frac{2v}{1+v}\right)\\
&+4\left(\frac{1}{v}\mathrm{Tanh}^{-1}v-1\right)\left(\frac{k_M}{3E_e}-\frac{3}{2}+\log\frac{2k_M}{m_e}\right)
+\frac{1}{v}\mathrm{Tanh}^{-1}v\left(2(1+v^2)+\frac{k_M^2}{6E_e^2}-4\mathrm{Tanh}^{-1}v\right)\biggr\}\biggr]. \nonumber
\end{split} \end{align}

\subsection{$I_1 \sim I_6$}\label{App-Ii}

    Using the Feynman parameter integral, we define $I_i$'s.
    Here, $I_1+I_3$ is derived from $\delta Z_p/2$.
    Similarly, $I_2+I_4$ is derived from $\delta Z_e/2$. 
    Also, $I_5+I_6$ corresponds to $iM_{1L}$.
    These are already canceled the ultraviolet divergence.
    The results are as follows.     
\begin{align} \begin{split}
I_1&=
\int_0^1 dx (1-x)\log[x^2m_p^2+(1-x) \mu^2]
=-\frac{3}{2}+\log m_p,\\
I_2&=
\int_0^1 dx (1-x)\log[x^2m_e^2+(1-x) \mu^2]
=-\frac{3}{2}+\log m_e,\\
I_3&=
\int_0^1 dx \frac{2x(1-x^2)m_p^2}{x^2 m_p^2+(1-x) \mu^2}
=-1+\log \frac{m_p^2}{\mu^2},\\
I_4&=
\int_0^1 dx \frac{2x(1-x^2)m_e^2}{x^2 m_e^2+(1-x) \mu^2}
=-1+\log \frac{m_e^2}{\mu^2}.
\end{split} \end{align}
\begin{align} \begin{split}
I_5=&
\int_0^1 dx  \int_0^{1-x} dy 
\frac{1}{\Delta-i\epsilon}\left\{(2-x)p'+y q'\right\}\left\{x p'+(2-y)q'\right\}
=I_{5a}+I_{5b}+I_{5c},
\end{split} \end{align}
    where $\Delta=m_p^2 x^2+m_e^2 y^2-(2x p'\cdot q'+\mu^2)y+(1-x)\mu^2$,
\begin{align} \begin{split}
I_{5a}
=&\int_0^1 dx  \int_0^{1-x} dy \frac{4p'\cdot q'}{\Delta-i\epsilon}
=
\frac{b}{c}\Bigl[\frac{4\pi^2}{3}+2i\pi\log\frac{2c}{c-c'}-\log\frac{2c}{c-c'}\log\frac{b+c}{b-c}+2Li(\frac{1+b-c}{1+b+c})\\
&+2Li(\frac{b-c}{b+c}\frac{1+b+c}{1+b-c})+\frac{1}{2}\log^2(\frac{1+b+c}{1+b-c})+\frac{1}{2}\log^2(\frac{b+c}{b-c}\frac{1+b-c}{1+b+c})\Bigr],\\
I_{5b}=&
\int_0^1 dx  \int_0^{1-x} dy \frac{2(p'-q')\cdot(xp'-yq')}{\Delta-i\epsilon}
=\frac{-1}{(1+b)^2-c^2}[(1+c^2-b^2)\log(b^2-c^2)-2c(\log\frac{b+c}{b-c}-2i\pi)],\\
I_{5c}
=&-\int_0^1 dx  \int_0^{1-x} dy \frac{(xp'-yq')^2}{\Delta-i\epsilon}
=
-\frac{1}{2}, \nonumber
\end{split} \end{align}
    where $b=p'\cdot q'/q'^2$, $c=\sqrt{(p'\cdot q')^2-p'^2 q'^2}/q'^2$, and $c'=\sqrt{(p'\cdot q')^2-p'^2 q'^2-(p'+q')^2\mu^2}/q'^2$,
\begin{align} \begin{split}
I_6
&=\int_0^1 dx  \int_0^{1-x} dy \bigl\{-2\log(\Delta-i\epsilon)\bigr\}
=3-\log m_e^2+\frac{1+b}{(1+b)^2-c^2}\log \frac{m_p^2}{m_e^2}
+\frac{2c}{(1+b)^2-c^2} 
\left(-\mathrm{Tanh}^{-1} \frac{c}{b}  +i\pi\right). \nonumber
\end{split} \end{align}
    We note here that $b/c=1/v$.

    The sum of these $I_i$'s is 
\begin{align} \begin{split}
\sum_{i=1}^6I_i
\simeq&
-\frac{5}{2}
+\log\frac{m_p^3 m_e }{\mu^4}
+\frac{2}{v}
\log\frac{\mu^2}{ m_e^2}\mathrm{Tanh}^{-1}v
-\frac{2}{v}Li(\frac{2v}{1+v})
-\frac{2}{v}(\mathrm{Tanh}^{-1}v)^2
+\frac{2\pi^2}{v}+\frac{2i\pi}{v}\log\left(\frac{4m_e^2}{\mu^2}\frac{v^2}{1-v^2}\right).
\nonumber
\end{split} \end{align}
    The last two terms diverge for $v\to 0$.
    The latter one, which contains $\mu$, does not affect the one loop decay width.
    However, this term has an nontrivial, important roll to two loop calculation. 


\end{document}